\documentclass[11pt]{article}
\usepackage{graphicx}

\newsavebox{\PSLASH}
\sbox{\PSLASH}{$p$\hspace{-1.8mm}/}

\begin{document}
\title{Loop models for CFTs }
\author{M. A. Rajabpour$^{a,b}$\footnote{e-mail: Rajabpour@to.infn.it} \\ \\
$^{a}$Institute for Studies in Theoretical Physics and Mathematics,
Tehran 19395-5531, Iran \\  $^{b}$Dip. di Fisica Teorica and INFN,
Universit{\`a} di Torino, Via P. Giuria 1, 10125 Torino,
\\Italy} \maketitle
\begin{abstract}
By interpreting the  fusion matrix as an adjacency matrix we
associate a loop model to every primary operator of a generic
conformal field theory. The weight of these loop models is given by
the quantum dimension of the corresponding primary operator. Using
the known results for the $O(n)$ models we establish  a relationship
between these models and  SLEs. The method is applied to  WZW, $c<1$
minimal conformal field theories  and other coset models.
  \vspace{5mm}%
\end{abstract}
\section{Introduction}\
The study of statistical mechanics systems related to loop models is
interesting both from the physical and the mathematical point of
views. Most of the statistical  models studied in physics, from the
Ising and the  q-state Potts model to more complex  vertex models,
can be represented in terms of loops. The loop representation of the
Ising model is very easy to understand: loops correspond to  domain
walls separating regions of opposite magnetization. If the Ising
model is defined on the  infinite plane or in a box with periodic
boundary conditions, then the boundary of domain walls  do not have
open ends or branch points. The same property  is true in  the more
general Potts model if we define domain walls appropriately.

A fundamental model, starting point in the  construction of more
complicated loop systems, is the $O(n)$ model and its connection
with exactly solvable models  led to the introduction of powerful
integrable model technologies. A first interesting progress in this
direction was the discovering of a critical point in the  $O(n)$
model made by Nienhuis \cite{Nienhis}, however the rigorous proof of
this result is still missing.

 After the advent of conformal field theory (CFT) as a
tool to classify  two dimensional critical phenomena,  the main
efforts to understand critical loop models have been based on the
CFT approach. Many critical loop models were defined and studied
also with other methods, such as the  Yang-Baxter
equation~\cite{Nienhis2} and non rigorous coulomb gas
methods~\cite{Nienhis3}. For a review of the coulomb gas method for
more complex loop models see \cite{kondev}. The critical points of
these models are described by CFTs, the early versions correspond to
well-known CFTs such as the  $c<1$ minimal models, but most of
recent proposals links to  CFTs with extra symmetries like the
Wess-Zumino-Witten (WZW) models~\cite{Di Franchesco}. For  recent
progress in this direction see~\cite{Fendley,Fendley2}. In parallel
with the progress on CFTs as a purely algebraic method to classify
critical points of statistical systems Pasquier in~\cite{Pasquier}
proposed the  ADE lattice models, a generalization of   the
restricted solid-on-solid models of Andrews, Baxter and Forester
\cite{ABF}, as the physical candidates for particular CFTs.
Pasquier's work is closely related to the Cappelli-Itzykson-Zuber
  ADE classification of CFTs~\cite{CIZ}   established by noticing  that
the operator content of many CFTs is constrained by  the modular
invariance of the partition function on  torus geometry and it is
related to Coxeter diagrams.

The definition of  the corresponding loop models and the study  of
the boundary changing operators for the $A$ series was initiated by
Saleur and Bauer in
\cite{SB}. The $A$ series is  related to dense critical loop models, while  the dilute
versions  were studied later by Kostov \cite{kostov}.

Some progress has been made in the  study  of loops related to the
remaining ADE lattice systems, but  many of their properties are
still unknown and  in particular the general fused ADE models are
not well explored. There are at  least a couple of reasons that make
the study of loop models related to statistical systems interesting.
Firstly, from the mathematical point of view it may give us  good
candidates for the Schramm Loewner evolution (SLE), a method
discovered by Schramm \cite{schramm} to  classify   conformally
invariant curves connecting two distinct boundary points in  a
simply connected domain. The parameter describing the curve is the
drift parameter $\kappa$ (For recent reviews see
\cite{cardy0,BB0}.). The exact relation between SLEs and CFTs was
discovered by Bauer and Bernard \cite{Bauer and Bernard1}:  using
the method of  null vectors of CFT they were able to  find a simple
relation between the CFT central charge and the drift $\kappa$. This
method, based on the original results of Schramm, was generalized
later to  loops without open ends by Sheffield and Werner
\cite{sheffield} and named Conformal Loop Ensemble (CLE).
 The connection between ADE loops  and CLE was studied by Cardy
\cite{Cardy ADE} in both dilute and dense cases by mapping the
height variable of ADE lattice models to the $O(n)$ model.

Another reason of interest is related to the study of topological
quantum field theory in $2+1$ dimensions and to its application to
topological quantum computation \cite{kitaev0}. The ground-state
wave function of these topological models coincides with the loop
ensemble. This means that the ground-state correlators in $2+1$
dimensional  topological quantum field theories are equal to
particular correlators  in 2D classical loop models \cite{freedman}.
The corresponding loops have weights related to the quantum
dimension of the operators appearing in the model. It seems that
mathematically one can classify the ground-state of topological
theories by two dimensional critical loop models or equivalently by
CFTs.

The above considerations suggest that a classification of the loop
representations related to  CFTs may be relevant for a  wide range
of applications.

In the following, using the  methods introduced by Cardy in
\cite{Cardy ADE}, we shall define a loop representation for a
generic CFT. To define these loop models we use the fusion matrix of
CFTs as an adjacency matrix and  map the resulting height model on a
$O(n)$ system. We find a different $O(n)$ model for every CFT
primary operator and observe that the weights of the loops coincide
with the quantum dimensions of the corresponding primary operator.
Using the link between  $O(n)$ and  CLEs  one can map every  CFT to
 SLEs and  to CLEs.

This allows the  investigation of the  link between SLEs and CFTs to
be carried out without the use of, the sometimes quite complicated,
null vector method. The main ingredient of the method is the  $S$
matrix  or the complete form of fusion matrix which is complex for
some  extended CFTs.

The paper is organized as follows: in section~2 we use the fusion
matrix as an adjacency matrix and define a height model using the
$S$ matrix as a weight for the plaquettes in a  triangular lattice.

Following \cite{Cardy ADE}, using the height model we will associate
to every height configuration an ensemble of loops, and by the
Verlinde formula  find that the  partition function of the loop
ensemble matches  that of a particular   $O(n)$ model.

In  section~3 we investigate the loop representation of some simple
CFTs such as the Ising and the  tricritical Ising models, the $c<1$
minimal models, the WZW  and in particular the $SU_{k}(2)$ models.
Our prediction for the  SLE drift parameter $\kappa$ is in agreement
with all the previously known examples.

In section~4 we will generalize  the methods introduced in section~2
and define loop models on the square lattice related to the dense
phase of the $O(n)$ models. The definition of loops  on the square
lattice is more natural because the  models arising from our
definition are always critical  without the need of a fine-tuning of
the parameters. In section~5 we will briefly discuss the connection
between  these  results and boundary CFT and show that the models
are well defined also in presence of a boundary.

Section~6  contains our conclusions with a  brief description of the
work in progress  motivated by  these results.

\section{ Loop models for general CFTs}\

To define loop models for a  given CFT we shall use  the modular $S$
matrix. To define the $S$ matrix we first need the expression of the
modular invariant partition function  on the torus geometry in terms
of  the characters and the parameter $\tau = e^{\frac{2\pi
i\beta}{R}}$, where  $R$ is the space length of the system and
$\beta$ is the time period, as follows
\begin{eqnarray}\label{Partition function}
Z=\sum M_{h,\bar{h}}
\chi_{h}(\tau)\chi_{\bar{h}}(\overline{\tau}).
\end{eqnarray}
In (\ref{Partition function}) the sum is over the operator content
of the theory and $M_{h,\bar{h}}$ is the non-negative integer that
specifies how many times a given representation enters, that is the
number of primary operators with scaling dimensions $(h,\bar{h})$.
The function $\chi_{h}=tr(\tau^{L_{0}-\frac{c}{24}})$ is the
character of the theory with the sum over the descendants of the
primary field with conformal weight $h$.

From now on  we shall use  $a,b,c$ to label the primary operators
(they are not the weights of primary operators) in the theory
independent of the holomorphicity of the operator. Then one can
define the modular $S$ matrix by  the modular transformation which
does not change the partition function but acts nontrivially on  the
characters
\begin{eqnarray}\label{S matrix}
\chi_{a}\left(-\frac{1}{\tau} \right)=\sum_{b}S_{a}^{b}\chi_{b}(\tau).
\end{eqnarray}
Another important ingredient is the fusion rule related to the
operator product expansion of a pair of primary operators,
\begin{eqnarray}\label{fusion rules}
\phi_{a}. \phi_{b}=\sum_{c}N_{ab}^{c} \; \phi_{c}~.
\end{eqnarray}
In (\ref{fusion rules}) $N_{ab}^{c}$ is the fusion coefficient and
the RHS involves   the entire tower of descendants of the primary
operator $\phi_{c}$. It is worthwhile to mention that it is possible
to interpret  the fusion coefficients as the elements of a matrix
$(N_{a})_{b}^{c}=N_{ab}^{c}$, so we have a fusion matrix for every
primary operator.\\
\begin{figure}
\begin{center}
\includegraphics[angle=0,scale=0.6]{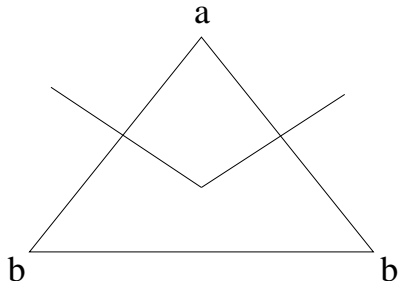}
\caption{A triangular plaquette  with $a\neq b$ and the
corresponding curve segment on the dual honeycomb lattice.}
\label{Fig1}
 \end{center}
\end{figure}
To define a loop model we  interpret the fusion matrix as the
adjacency matrix associated to a graph~\cite{Cardy ADE}. The graph
of a primary operator $\phi_{a}$ has $g$ vertices where $g$ is the
number of primary operators in the theory and edges connecting pairs
of vertices  $(b,c)$ when $N_{ab}^{c}=1$. Following \cite{Cardy ADE}
one can define a height model on the triangular lattice by imposing
that the height $h_{j}$ at the site $j$ can take values
$0,1,\dots,g-1$. Then constraint the heights at neighboring sites
according to the incidence matrix  associated to a given primary
field $\phi_a$: only neighbor heights $h_{k}$ and $h_{k}$ with
$(N_{a})_{h_j}^{h_k}=1$ are admissible. One should notice that
according to the fusion matrix rule  neighbor heights  maybe
identical, which is not the case for ADE models. Actually, for a
consistent definition of loop models on a triangular lattice at
least two of the heights at the corners of  an elementary triangular
plaquette should be equal \footnote{ We can have  loops in our graph
but at least for minimal models and restricting the analysis to
primary operators of  interest, it is possible to prove that there
are not graphs with loops smaller or equal to three. In general  we
just define models for graphs with loops longer  than three  or even
without loops. } then the weights for the elementary plaquette are
defined  as follows. If the  heights of plaquette are $(c,b,b)$ with
$c \ne b$ then weight is $x(\frac{S_{k}^{b}}{S_{k}^{c}})^{1/6}$,
where $S$ is the modular matrix and $k$ is arbitrary. If the heights
are all equal then the weight is $1$ except for those with
$N_{ab}^{b}\neq 0$ that have  weights  $1$ or $x$ depending on the
particular model considered, as it will be explained below in the
paper.

The next step is to mark triangles with unequal heights $(c,b,b)$
drawing a curved segment on  the dual honeycomb lattice \cite{Cardy
ADE} and linking to the center the midpoints of the two  edges with
different  heights ($b$ and $c$) at the extremes (See figure 1). The
difference with the more standard  ADE models is related to the
equal height $(b,b,b)$ plaquette with $N_{ab}^{b}= 1$. For
plaquettes with height $(b,b,b)$ there are three possibilities for
drawing curve segments. In these cases we will suppose, as in
percolation problems, that the lines in dual honeycomb lattice
choose randomly two of the edges of triangular lattice consistently
with the other sites. This define a loop configuration for every
height configuration. Summing over the admissible values of heights
consistent with a given loop configuration we find
\begin{eqnarray}\label{loop height}
\sum_{b}(N_{a})_{b}^{c}\frac{S_{k}^{b}}{S_{k}^{c}}=\frac{S_{k}^{a}}{S_{0}^{k}},
\end{eqnarray}
where the sum is just over $b$. To get this formula we have used the
Verlinde formula for CFTs,
$\sum_{b}(N_{a})_{b}^{c}\frac{S_{k}^{b}}{S_{k}^{0}}=\frac{S_{k}^{a}}{S_{0}^{k}}\frac{S_{k}^{c}}{S_{k}^{0}}$,
which means that the $b$th element of the eigenvector of $N_{a}$
with eigenvalue $\frac{S_{k}^{a}}{S_{0}^{k}}$ is given by
$\frac{S_{k}^{b}}{S_{k}^{0}}$. We  take always $k=0$ to get the
largest eigenvalue and have positive real weights in our height
models. For properly-defined fusion matrices these eigenvalues will
coincide with the quantum dimension of $\phi_{a}$:
\begin{eqnarray}
\label{quantum dimension}
d_{a}=\frac{S_{0}^{a}}{S_{0}^{0}}.
\end{eqnarray}
 Summing iteratively over all heights
of all  clusters gives a factor $d_{a}$ for each closed loop, so the
partition function of our model has a $O(n)$ like partition function
\begin{eqnarray}\label{O(n)}
Z=\sum x^{l}d_{a}^{N},
\end{eqnarray}
where $l$ is the number of bonds in the loop configuration and $N$
is the number of loops.

Using this  method we are  then able to associate  to every primary
field of a given CFT an $O(n)$ model with $n=d_{a}$. Let us discuss
some simple but general properties of these models. Firstly it is
not difficult to see that for the  identity operator our incidence
graph has $g$ disconnected points with a blob attached to them to
indicate that they are adjacent to themselves. This means that the
configuration space is decomposable and  that all  the heights
should be equal. In loop language means that we  have a percolation
or an Ising loop model depending on the value of $x$. It is not
difficult to see that for any non-connected graph our height
configuration space is decomposable and that we can treat the
different parts of our graph separately. Another interesting fact is
related to the consistency of different parts of graph: if we write
for different connected components of graph the corresponding
adjacency matrices then all the parts will have the same loop
weights and the same quantum dimension. Therefore,  we can
consistently keep only one part of the graph and find the loop model
associated to it. We shall clarify further this issue later  by
discussing some minimal model examples in details.

It was found long time ago in~\cite{Nienhis} that the $O(n)$ model
posses a dilute critical point for $n\leq 2$ with
$x_{c}=\frac{1}{\sqrt{2+\sqrt{2-n}}}$: correspondingly  our loop
models will have a critical point just for the fields with quantum
dimension smaller than $2$. The $O(n)$ model  has another critical
regime, the so-called dense phase, for $x=(x_{c},\infty)$ which
corresponds to a different universality class. The mapping to the
$O(n)$ model helps us to find the connection with CLE. From coulomb
gas arguments we know that, in the dilute regime, the loop weight
has the following relation with the drift in the CLE equation,
\begin{eqnarray}\label{S minimal}
d_{a}=-2 \cos(\frac{4\pi}{\kappa}).
\end{eqnarray}
We will use this equation to establish a connection between our
models and CLEs and find  that the  prediction is  in excellent
agreement with previous results. We should mention that for the
dense phase the above equation is still true if we work in the
region $4\leq \kappa \leq 8$. In the next section we will
investigate the loop representation of some simple CFTs.

\section{Some examples}\
 In this section we show how the method described in the
previous section is working for some simple models such as minimal
models, $SU_{k}(N)$ WZW models and coset models.
\subsection{Minimal Models:}
\setcounter{equation}{0}
 Consider the
minimal models $M(p',p)$, the simplest case is the Ising model
$M(4,3)$. This model has three primary operators \cite{Di
Franchesco}: $1,\sigma,\epsilon$ with  graphs depicted in figure~2.
We discussed in the previous section how the operator $1$ is related
to percolation or Ising model with $\kappa=6$ or $3$. The other two
cases are more interesting, for example the graph of $N_{\epsilon}$
has the part similar to the $A_{2}$ graph and a disconnected point,
using $d_{\epsilon}=-2 \cos(\pi\frac{4}{3})=1$, it is not difficult
to see that both parts have the same dominant eigenvalues
corresponding to $\kappa=3$ which is in agreement with \cite{Cardy
ADE}. The graph for $N_{\sigma}$ is similar to $A_{3}$ and gives us
$\kappa=\frac{16}{3}$ in the dense phase, which is the dual of the
previous case. It is quite surprising that the critical behavior of
these loops are completely in agreement with the known Ising model.
\begin{figure}
\begin{center}
\includegraphics[angle=0,scale=0.8]{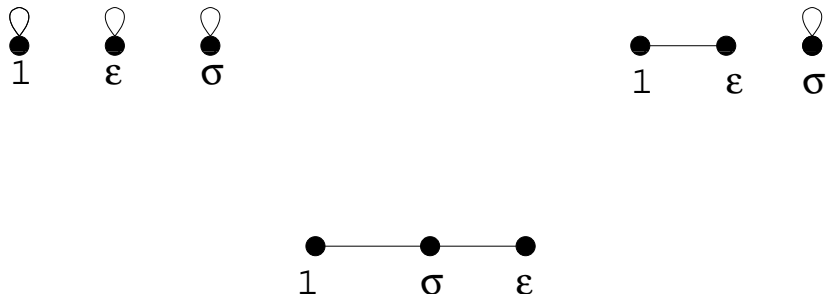}\\
\caption{ Graphs of fusion matrices of primary operators in Ising
model, from left to right the fusion graph of $1$, $\epsilon$ and
$\sigma$. } \label{Fig2}
 \end{center}
\end{figure}

The next simple minimal model is tricritical Ising model $M(5,4)$
with six primary operators and fifteen different fusion relations
\cite{Di Franchesco}. The fusion matrix graphs for this model are
shown in the Fig~3. The familiar cases with the critical loop models
are $1, \sigma', \epsilon$. The graph of $N_{\sigma'}$ has two
$A_{3}$ parts, using $d_{\sigma'}=-2 \cos(\pi\frac{5}{4})$ it is
easy to get $\kappa=\frac{16}{5}$ which is again in agreement with
\cite{Cardy ADE}. The case $N_{\epsilon}$ is similar to $A_{4}$ and
related to $\kappa=5$. One can check simply that the other part of
the graph has the same eigenvalue and so related to the same kind of
loop model, this kind of graphs are the first kind of non-familiar
height models corresponding to loop models. We also notice that one
part of this graph is similar to the second level of the fusion of
$A_{4}$ model. We have two more graphs related to critical loop
models, $N_{\epsilon'}$ has quantum dimension equal to the
$N_{\epsilon}$ so they have the same properties. The quantum
dimension related to $N_{\epsilon''}$ is equal to one and so again
we have percolation or Ising model. The last one is the $N_{\sigma}$
which has quantum dimension bigger than two and so we can not
associate a critical loop model to this primary operator, we notice
that this graph is also similar to ground sate adjacency diagram of
the second level fusion of $A_{6}$ model ~\cite{Fendley}.

Let's summarize the meaning of disconnectedness in the graphs of
different fusion matrices of primary operators. For both of the
above models if we treat the different parts of graph as the
different models then we can find finally the same loop models. This
means that the different blocks of our block diagonalizable
adjacency matrices have the same largest eigenvalues and here our
definition of loop model is consistent. One can always choose one
block and get the corresponding loop model. This is exactly the same
as the definition of the well known ADE models, for these models
always getting one block of our fusion matrix is enough. The
degeneracy in the eigenvalues of the fusion matrix is the natural
result comes from the most general property of fusion matrix:
different fusion matrices commute with themselves. It seems that the
above argument is a general property of many different CFTs
specially the more general cases which we will investigate in the
following.
\begin{figure}
\begin{center}
\includegraphics[angle=0,scale=0.6]{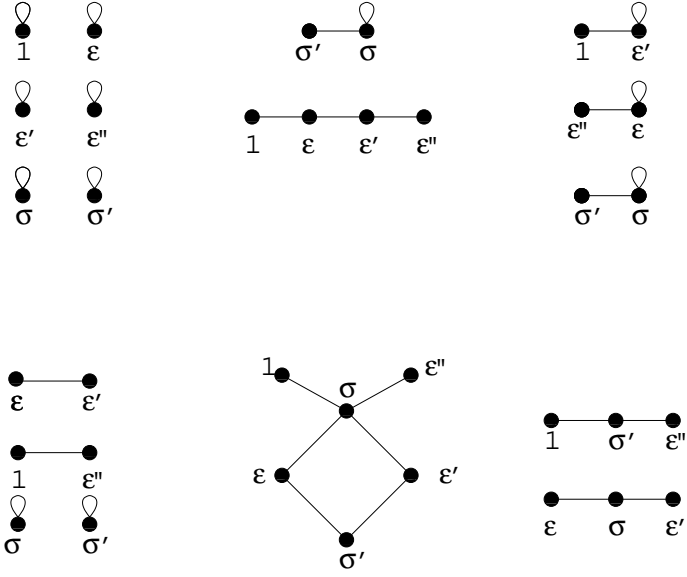}\\
\caption{Graphs of fusion matrices of primary operators in three
critical Ising  model in the upper row from left to right the fusion
graph of $1$, $\epsilon$ and $\epsilon'$, in the lower row from left
to right the fusion graph of $\epsilon''$, $\sigma$ and $\sigma'$.}
\label{Fig3}
 \end{center}
\end{figure}

The above calculation is tractable in the general case of minimal
models $M(p',p)$ with Kac formula for the conformal weights of
primary operators
\begin{eqnarray}\label{Kac formula}
h_{r,s}=\frac{(p'r-ps)^{2}-(p-p')^{2}}{4pp'},
\end{eqnarray}
where (rs) is the label of primary operators in the Kac table and
$1\leq r \leq p-1$, $1\leq s \leq p'-1$. For the known primary
operators in the non-extended Kac table, it is not difficult to see
that because of $(r,s)=(p-r,p'-s)$ there is a huge redundancy in
this rectangle. The $S$ matrix for minimal models has the following
form

\begin{eqnarray}\label{S minimal}
S_{rs,\rho
\sigma}=2\sqrt{\frac{2}{pp'}}(-1)^{1+s\rho+r\sigma}\sin(\pi\frac{p}{p'}r\rho)\sin(\pi\frac{p'}{p}s\sigma).
\end{eqnarray}
Using equation (\ref{quantum dimension}) the quantum dimension of
 primary operator $\phi_{rs}$ has the following form
\begin{eqnarray}\label{Quantum dimension minimal}
d_{rs}=(-1)^{r+s}\frac{\sin(\pi\frac{p}{p'}r)\sin(\pi\frac{p'}{p}s)}{\sin(\pi\frac{p}{p'})\sin(\pi\frac{p'}{p})}.
\end{eqnarray}
We should mention that the above equation respects the symmetry
$(r,s)=(p-r,p'-s)$ for the unitary minimal models. One can
investigate the properties of graphs for different operators but the
most interesting cases are $(r,s)=(1,2)$ and $(r,s)=(2,1)$. The
fusion matrices for these cases are the following symmetric
matrices,
\begin{eqnarray}\label{fusion matrix minimal}
N_{(1,2)(r,s)}^{(m,n)}=\delta_{m,r}(\delta_{n,s+1}+\delta_{n,s-1})
\hspace{2cm}
N_{(2,1)(r,s)}^{(m,n)}=\delta_{n,s}(\delta_{m,r+1}+\delta_{m,r-1}).
\end{eqnarray}
It is not difficult to see that the graph for these two cases do not
have any loop and so we do not need to worry about possibility of
definition of loop model. For example one can show that the graph of
$(r,s)=(1,2)$ for unitary minimal models, $M(p+1,p)$, has always a
part similar to the graph $A_{p-1}$ and $(r,s)=(2,1)$ has a part
similar to $A_{p}$.  The quantum dimension of $(r,s)=(1,2)$ is
\begin{eqnarray}\label{Quantum dimension 12}
d_{12}=-2\cos(\pi\frac{p'}{p}).
\end{eqnarray}
If we suppose that the loop model is at the critical regime then it
is easy to see that this loop is related to CLE with the following
drift,
\begin{eqnarray}\label{k minimal}
\kappa=4\frac{p}{p'},
\end{eqnarray}
which is exactly the same as the $\kappa$ predicted for minimal
models by using null vectors \cite{Bauer and Bernard1}. One can
see this by rewriting the level $2$ null vector of minimal models
as follows,
\begin{eqnarray}\label{2 null vector}
L_{-2}-\frac{3}{2(2h+1)}L_{-1}^{2}=L_{-2}-\frac{p}{p'}L_{-1}^{2},
\end{eqnarray}
for the unitary Virasoro minimal models and comparing it with the
null vector like relation in SLE \cite{Bauer and Bernard1}. It seems
that there is a close relation between this loop model for
$(r,s)=(1,2)$ and CLE. The same calculation is possible for
$(r,s)=(2,1)$ and the result is $d_{21}=-2 \cos(\pi\frac{p}{p'})$
with $\kappa=4\frac{p'}{p}$ which is apparently the dual of the
previous case and consistent with CLE prediction\footnote{ The outer
boundary of fractal curve with $\kappa>4$ is given by an other
fractal curve with $\kappa:=16/\kappa$ named the dual curve. This
curve has the same central charge as the original one.}. In this
construction it is also possible to get other CLE drifts by
considering the dense (dilute) phase of the loop model corresponding
to $\phi_{1,2}$ ($\phi_{2,1}$). In this level it is not clear how
one can choose the right phase but the algorithm for minimal models
as we noticed is taking dilute phase for $\phi_{1,2}$ and dense
phase for $\phi_{2,1}$. For exact correspondence one need to
investigate the problem in the partition function level.

These two cases are the familiar cases but as we noticed in
tricritical Ising model these are not the only loop models with
critical point, in fact it is possible to have many critical loop
models for a minimal model with generic $p$ and $p'$ just we need to
have quantum dimension smaller or equal to two. Appearing the
$A_{p-1}$ and $A_{p'-1}$ graphs in our fusion matrix is not just an
accident because in fact these are the graphs that already appeared
in the ADE classification of minimal models \cite{CIZ}. From the
other point we know that these graphs have lattice statistical
physics counterparts with the loop representations investigated in
\cite{Cardy ADE}, so appearing a close connection between
classification of modular invariant CFTs in two dimensions with
respect to the ADE graphs with our loop representation is not so
much surprising. At least for the minimal models this connection is
obvious and the fusion graphs of some special operators are exactly
the same as the ADE diagram of the corresponding CFT.

For later purposes let's write the results explicitly for the case
$M(6,5)$ which is describing the three state Potts model, using
equation (\ref{Quantum dimension minimal}) we find the following
quantum dimensions,
\begin{eqnarray}\label{m(6,5)}
d_{(1,1)}&=&1,\hspace{1.7cm}d_{(1,2)}=2\cos(\frac{\pi}{5}),\hspace{1.2cm}d_{(1,3)}=2\cos(\frac{\pi}{5}),\hspace{1cm}d_{(1,4)}=1,\hspace{1cm}\nonumber\\
d_{(2,1)}&=&\sqrt{3},\hspace{1.3cm}d_{(3,1)}=2,\hspace{2.4cm}d_{(2,2)}=2\sqrt{3}\cos(\frac{\pi}{5}),\hspace{0.5cm}d_{(2,3)}=2\sqrt{3}cos(\frac{\pi}{5}),\hspace{1cm}\nonumber\\
d_{(3,3)}&=&4\cos(\frac{\pi}{5}),\hspace{0.4cm}d_{(3,2)}=4\cos(\frac{\pi}{5}),\hspace{1.25cm}d_{(2,4)}=\sqrt{3},\hspace{1.9cm}d_{(3,4)}=-2.
\end{eqnarray}
It is evident that different primary fields can have the same
quantum dimension and so the same loop representations. For example
$\kappa_{1,2}=\kappa_{1,3}=\frac{10}{3}$ and
$\kappa_{2,1}=\kappa_{2,4}=\frac{24}{5}$, however we know from the
connection between CFT and SLE that the fields $(1,2)$ and $(2,1)$
are connected to SLE by null vector so we need to answer this
question how we can choose the right operator. For three state Potts
model we can answer the question by looking at the modular
invariancy of the model which is responsible for constraining the
operator content of the model to just six operators, $(1,1)$,
$(2,1)$, $(3,1)$, $(4,1)$, $(3,3)$, $(1,3)$ in the spin
representation of the model \cite{Di Franchesco}. However for
Fortuin-Kasteleyn representation of the model this operator content
is not enough and we should insert disorder operators as well. The
theory has extra $W$ symmetry which is responsible in decreasing the
primary operators of the theory from ten to six. If we look at the
model with the above symmetry then we can show that this model is
not just a sub theory of minimal model $M(6,5)$, the fields in this
model represent different primary fields with different characters
however the conformal dimensions are equal. Using operators $W(z)$
and $\overline{W}(\bar{z})$, with weight equal to three, one can
show that $\phi_{(1,3)}=W_{-1}\overline{W_{-1}}\phi_{(1,2)}$,
$\phi_{(2,3)}=W_{-1/2}\phi_{(2,2)}$ and
$\phi_{(4,1)}=W_{-1/2}\phi_{(2,1)}$, where $W_{n}$ and
$\overline{W}_{n}$; $n=0,\pm 1,  \pm2,...$ are the mode operators.
The above relations show that the operators with the same quantum
dimension are in fact related with the $W$ symmetry of the model. In
the other word they are appearing in the modular partition function
together \cite{Di Franchesco}, so the equality of quantum dimension
of different fields are a sign for the internal extra symmetry of
the model. This interpretation of three states Potts model is also
consistent with the parafermionic interpretation
\cite{Zamolodchikov}. The same story is true for tricritical Ising
model which has also supersymmetry at the critical point. This
symmetry is responsible for the equality of quantum dimension of the
operators $\textbf{I}$ and $\epsilon''$ and also the operators
$\epsilon$ and $\epsilon'$. Both couples come from Neveu-Schwarz
sector of superconformal symmetry of the model \cite{Di Franchesco}
and one is the descendent of the other in the presence of
supersymmetry. The other important thing to mention is to get the
connection to the physical system we need to investigate the
symmetries of the statistical model specially those that are
connected to the symmetries of the domain walls. For example for
three states Potts model the fluctuating domain walls should respect
the $Z_{3}$ symmetry. We know from the boundary conformal operators
that the corresponding responsible operator is the spin operator
\cite{cardy1989}, but
 free boundary condition is not respecting $Z_{3}$ symmetry and so in the modular
 invariant partition function we do not have the corresponding operator, so we always need to have some informations coming
from the boundary CFT as well.

 The above calculation can be generalized for
the extended Kac table too, for example take the simple case
$M(3,2)$ which is related to CFT with zero central charge. The loop
model for the operator $(1,2)$ has $\kappa=\frac{8}{3}$ which is
self avoiding random walk with zero quantum dimension. The loop
model of $(2,1)$ has $\kappa=6$ which is percolation. These two
cases are again consistent with the known results. We should mention
that the above naive argument is just coming from equation (\ref{k
minimal}) and does not mean that the adjacency matrix is just
$A_{1}$ . The other simple case is $M(2,1)$ with quantum dimension
$-2$. In this case we have $\kappa=2$ and $\kappa=8$, corresponding
to loop erased random walk and spanning trees respectively. One can
simply investigate the extended Kac table of more complex CFTs like
Ising model and find some new loop models with zero or negative
quantum dimension which is related to $O(n)$ models. For example in
Ising model $(1,4)$ has zero quantum dimension and $(1,5)$ has
negative quantum dimension. The zero and negative quantum dimensions
can not be related to unitary CFTs and as it is quite well known the
extended Kac table models are not unitary, they are related to
logarithmic CFTs (LCFT) with special kinds of fusion matrices. For
example let's briefly see the most familiar LCFT, $c=-2$. For this
case if we just suppose the fields with
$(r,s)=(1,1),(1,2),(1,3),(1,4),(1,5)$ then the fusions of the
operators make a closed algebra \cite{Flohr}. Using the fusion
matrices of operator $(1,2)$, which has also integer elements more
than one, will give us the same dominant eigenvalue equal to zero,
this is the same as our above naive argument. We think that these
loop models are in close relation with the lattice logarithmic
minimal models introduced by Pearce, Rasmussen and Zuber \cite{prz}.
\newline
\newline
\subsection{ $SU_{k}(N)$:}

Let's investigate the CFTs with additional symmetries specially WZW
models and for simplicity the most simple and familiar case i.e,
$SU_{k}(2)$. The fusion matrix of this models are well known; for
example see \cite{Di Franchesco}, by generating loop model as before
we will have some loops with weights equal to the quantum dimension
of the corresponding primary operators, for this case quantum
dimensions are completely familiar and has the following form

\begin{eqnarray}\label{Su quantum dimension}
d_{a}=\frac{\sin(\pi(a+1)/(k+2))}{\sin(\pi/(k+2))},
\end{eqnarray}
where $a=2j$ with $j=0,1/2,...,k/2$ is the spin of the
representation. The only nontrivial case with quantum dimension
smaller than $2$ is related to $a=1$, the spin $1/2$, which has
quantum dimension $d_{1}= 2 \cos(\frac{\pi}{k+2})$. The CLE drift is
\begin{eqnarray}\label{SU}
\kappa=4\frac{k+2}{k+3}
\end{eqnarray}
the same as the \cite{Gruzberg} result which was found by using null
vector relations. We notice that just for $k=4$ we have another
quantum dimension which is equal to $2$ and related to $a=2$, spin
$1$, the corresponding CLE drift is  $\kappa=4$. We notice that this
result is the same as the result coming from $A_{k+1}$ with the same
adjacency graph and so in the close connection with the modular
invariant partition functions of this model. The other example is
$SU_{k}(3)$ with the quantum dimensions,
\begin{eqnarray}\label{Su3 quantum dimension}
d_{a,b}=\frac{\sin(\pi(a+1)/(k+3))\sin(\pi(b+1)/(k+3))\sin(\pi(a+b+2)/(k+3))}{\sin(\pi/(k+3))\sin(\pi/(k+3))\sin(2\pi/(k+3))},
\end{eqnarray}
where $a,b\geq0$ and $a+b\leq k$. For the simplest case $k=1$ we
have two quantum dimensions $d_{1,0}=1$ and $d_{0,1}=1$, for $k=2$
we have three different quantum dimensions
$d_{1,0}=d_{0,1}=d_{1,1}=2\cos(\frac{\pi}{5})$, $d_{2,0}=d_{0,2}=1$.
It seems that there is not a nice compact form for the general case
but the calculation for larger $k$s is straightforward and shows
that most of the quantum dimensions are greater than two, except
some operators with quantum dimensions equal to one. The same
calculation for the $SU_{k}(N)$ is tractable and for $k=2$ one of
the critical loop models has the following CLE drift,
\begin{eqnarray}\label{SUN}
\kappa=4\frac{N+2}{N+3}.
\end{eqnarray}
This is similar to equation (\ref{SU}) if we replace $k$ with $N$,
this is a general property in $SU_{k}(N)$ models and reminiscent of
level-rank duality. One can find the quantum dimension of the most
general affine
 Lie algebras as the weights of loop models and try to find the
 CLE drift by using $O(n)$ results, to find quantum dimensions of many CFT's see \cite{Di Franchesco}.

 \subsection{Coset Models}

  \hspace{0.8cm}$Z(k)$ model: The same calculation can be done for the $Z(k)$
  models
related to the coset, $\frac{SU_{k}(2)}{U(1)}$, with the same fusion
rules and operator content but with the different conformal weights.
The central charge of this model is $c=\frac{2(k-1)}{k+2}$ and the
conformal weights of the primary fields $\phi_{l,m}$ are given by
$\Delta_{l,m}=\frac{l(l+2)}{4(k+2)}-\frac{m^{2}}{4k}$ with $0\leq l
\leq k$, $0\leq |m|\leq 2k-1$ and $l-m \in 2\textbf{Z}$ so that we
should just take half of the grid to remove the redundancy, the spin
operator is corresponding to $m=l$ in this notation. The modular $S$
matrix are given explicitly by
\begin{eqnarray}\label{SZK}
S_{i,j}^{m,n}=\frac{1}{\sqrt{k(k+2)}}\sin\frac{\pi(i+1)(j+1)}{k+2}\exp
(i\frac{\pi jn}{k}.).
\end{eqnarray}
 Since these models have
the same fusion rules as the $SU_{k}(2)$ models we find the same
quantum dimensions and so re-derive the equation (\ref{SU}) for the
first spin operator of this model. For this case if we go to the
dense phase of the loop model then we have
\begin{eqnarray}\label{Z(k)}
\kappa=4\frac{k+2}{k+1},
\end{eqnarray}
 which is the same as the formula proposed in \cite{Santachiara} for the lattice $Z(k)$
 models in the FK representation.
 We were not able so far to find the formula for
 the spin representation of these models by the familiar $S$
 matrix of $Z(k)$ models. The
 $Z(2)$ case is the Ising model and it is equal to the minimal model $M(4,3)$ and so there is not
 any ambiguity in this case. For $Z(3)$ it is true that the model
 is equal to three states Potts model in the spin representation however the operator content
 of this model is not equal to $M(6,5)$. However the two disordered operators appearing in the Dihedral
 description of the $3$ state Potts model is missing in this description but  the equation (\ref{SU}) is giving true answer for $k=3$. For higher $k$'s the result is different from
  the result of \cite{Santachiara}, we will come back again to this problem when we shall discuss loop models on square lattice. The above argument means that we should be
 careful in extending our results to the loop models in the
 lattice $Z(k)$ model defined by spin variables.\\

 $\frac{SU_{k}(2)\oplus SU_{l}(2)}{SU_{k+l}(2)}$: The similarity of the quantum dimension of
 $\frac{SU_{k}(2)}{U(1)}$ and $SU_{k}(2)$ comes from the general property
 of coset models, however coset models can have the larger operator
 content with different conformal weights but they have most of the quantum dimensions of original
 affine algebra. For example take the coset,
 $\frac{SU_{k}(2)\oplus SU_{l}(2)}{SU_{k+l}(2)}$, for $l=1$ it is equal
 to $M(p+1,p)$ minimal models with $\kappa=4\frac{k+2}{k+3}$ for operator $(1,2)$ which
 is equal to the $SU_{k}(2)$ case however the $M(p+1,p)$ minimal
 models have other critical loop representations as well.
 Generalization of these results to more complicated coset models is
 straightforward and we discussed some of them in the Appendix.

\section{Loop Models on Square Lattice}\
\setcounter{equation}{0}

 The definition of loop models on square lattice is similar to the
 honeycomb lattice. From some point of views it is more
 natural, for example for the honeycomb lattice we introduced a
 non-known parameter $x$ which it should be fine tuned to get a
 critical loop model but on the square lattice just the $S$ matrix
 is enough to get a critical loop model. The definition is as
 follows: put some heights on the vertices of the lattice so that
 the neighbor sites be adjacent on the fusion matrix
 graph. As the honeycomb lattice if the fusion graph has no
 cycles smaller than five then with probability one, one of the
 diagonally opposite pairs of heights in each elementary plaquette
 should be equal, we will use this constraint for definition of
 loops. Labeling the heights by $(a,b,c,d)$, the weight around
 elementary plaquette is

\begin{eqnarray}\label{weight squire}
W(a,b,c,d)=(\frac{S_{b}S_{d}}{S_{a}S_{c}})^{^{\frac{1}{4}}}\delta_{ac}+\lambda(\frac{S_{a}S_{c}}{S_{b}S_{d}})^{^{\frac{1}{4}}}\delta_{bd},
\end{eqnarray}
where $\lambda=1$ for the unitary minimal models. So far the
definition is similar to the non-dilute ADE models but as we noticed
in the first section there are some fusion matrices with diagonal
terms. This means that it is possible to have some neighboring
vertices with the same heights, but still we will have at least a
diagonally opposite pairs of heights in each elementary plaquette
with the same height for most of the cases that we are interested.
For these cases we define the weight of the plaquette by using loop
model as follows: in each elementary plaquette we draw an edge
connecting the diagonal sites if they are equal then we will have a
Fortuin-Kasteleyn like graph \cite{Cardy ADE} and then the
definition of loop model on the medial lattice is straightforward
just as Pott's model, Fig~4. Each part of the loop inside plaquette
carries a weight, $(\frac{S_{b}}{S_{a}})^{^{\frac{1}{4}}}$.
\begin{figure}
\begin{center}
\includegraphics[angle=0,scale=0.6]{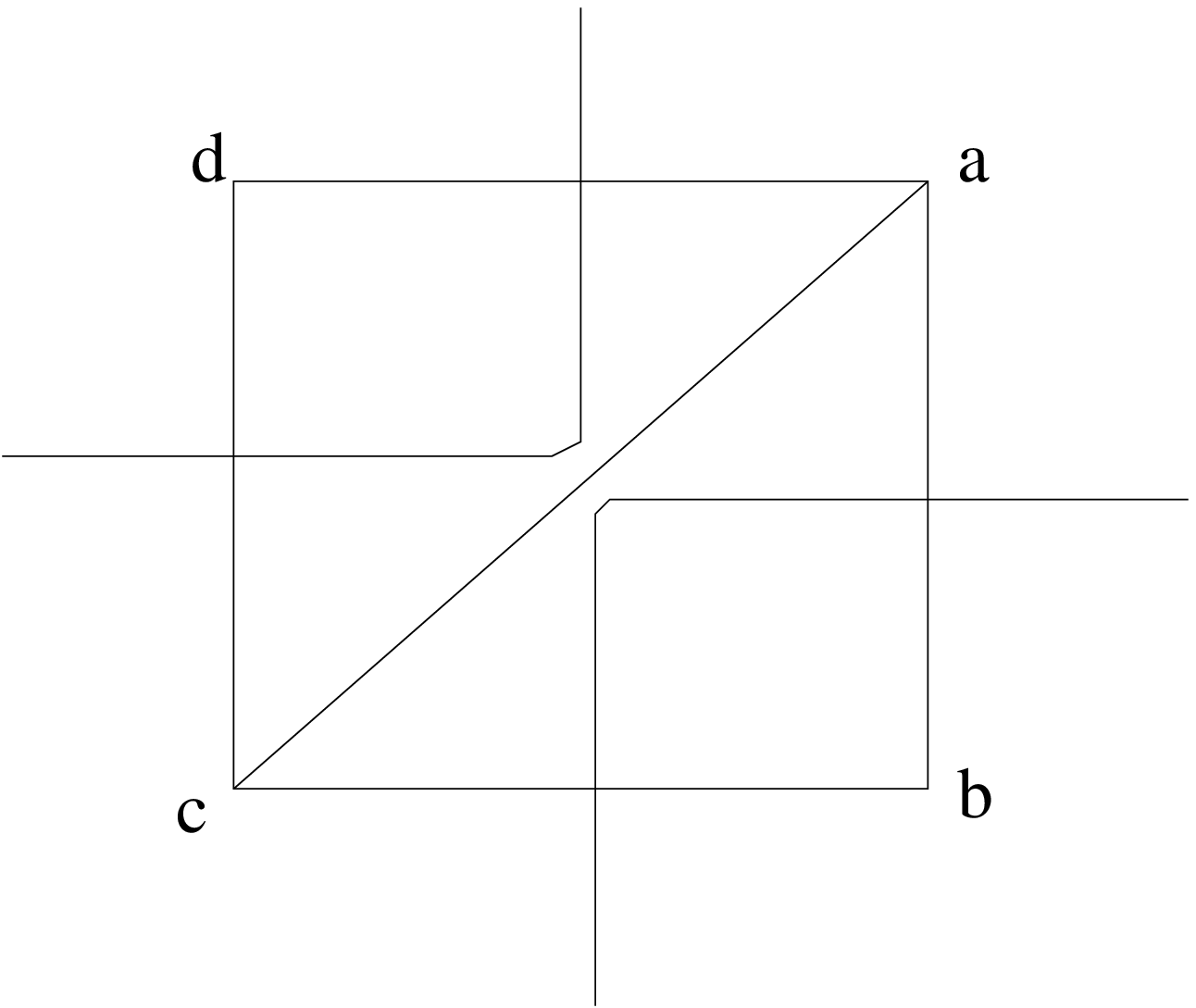}\\
\caption{A plaquette of square lattice with $a=c$ and the
corresponding curve segment on the dual square lattice}
\label{Fig.4}
 \end{center}
\end{figure}
For plaquette with three equal heights for drawing the edges we do
not have any ambiguity and we can use the equation (\ref{weight
squire}) to get the weight of the plaquette. For plaquette with four
equal heights we chose one of the edges like percolation with
weights equal to one and after that the definition of loop model is
straightforward. If we consider the weights of all parts of loop and
then sum over all values of the heights consistent with a
configuration of loops we will have an $O(n)$ model with $x=1$ which
is exactly the same as the loops in the FK representation of the
$Q=d_{a}^{2}$ state Potts model. Following \cite{riva cardy} there
should be some holomorphic operators with conformal weights equal to
$s$ satisfying the equation $d_{a}=2\sin(s\frac{\pi}{2})$. This
holomorphic operators describe the probability that the curve passes
between two close points. Then the corresponding drift term of CLE
is
\begin{eqnarray}\label{drift2}
\kappa=\frac{8}{1+s}.
\end{eqnarray}
The above argument shows that we can map all of the loop models
defined by the above method to the Potts model and then by using the
known results of critical Potts model we can get all of the
properties of loop models. Let's investigate some examples:
\newline
\newline
Example~1: Minimal models

For the case $(r,s)=(2,1)$ it is easy to see that the spin of
corresponding operator should be equal to $h_{3,1}=2\frac{p}{p'}-1$
and we have $\kappa=4\frac{p'}{p}$ consistent with the argument in
\cite{riva cardy}. One can find the results for the other cases by
using the same technics. In all of the cases the results are
consistent with the results of the previous section.
\newline
\newline
Example~2: Z(k) model

The calculation for the parafermionic models are the same and we can
find easily the connection of these CFTs to CLE by the above method.
One of the famous examples is Z(k) model with the following
connection to CLE for the operators with spin $\frac{1}{2}$
representation,

\begin{eqnarray}\label{drift3}
\kappa=4\frac{k+2}{k+1},
\end{eqnarray}
where the corresponding holomorphic operator describing the
probability that the curve passes between two close points is,
$s=\frac{k}{k+2}$. We conjecture that the above formula, as
Santachiara pointed out, could correspond to the FK representation
of lattice $Z(k)$ model with $k>3$. However the result is coming
again from the different operator, in \cite{Santachiara} the author
found the above equation by using one of the disordered operators
comes from dihedral group but we do not have these operators in our
calculations. This is another example for different primary
operators with the same quantum dimensions. We will face again with
this ambiguity when we discuss lattice $Z(4)$ model, Ashkin-Teller,
as an orbifold model \cite{rajabpour}. However there are some
evidences that $Z(k)$ parafermionic model is a good candidate for
the lattice $Z(k)$ model \cite{rajabpourcardy} but it seems that it
is not enough for describing all of the properties of the model. In
addition it shows that it is important to investigate the fusion
matrix of the dihedral group to improve our understanding about
lattice $Z(k)$ models. For $Z(k)$ parafermionic CFT there is another
more straightforward evidence that the formula (\ref{drift3}) is a
good candidate, the argument is as follows: In equation (\ref{weight
squire}) if we put $\lambda=-1$ then the corresponding CFT
describing the height model is $Z(k)$ model which is also
antiferromagnetic Potts model \cite{Fendley2}; see also
\cite{saleur}, and related to loop model with weight $-d_{a}$. then
again by following \cite{riva cardy} it is easy to show that there
is antiholomorphic operator with spin
$d_{a}=-2\sin(s\frac{\pi}{2})$, this antiholomorphic operators are
describing the probability that the curve passes between two close
points but in this time by the following connection to CLE drift
\begin{eqnarray}\label{drift3}
\kappa=\frac{8}{1-p},
\end{eqnarray}
which again gives the equation (\ref{drift3}). The tricky point is
we need to work with the antiholomorphic operator in this case.

To close this section we just should mention that the loop models
coming from the definition on the square lattice is just the dense
phase of loop models discussed in the previous sections, this is
easy to understand if we suppose $x=1$ in the $O(n)$ model's
partition function. This is always true because for $-2\leq n\leq
2$, $x_{c}$ is always smaller than one.

\section{Discussion: connection to BCFT}\

In this section we would like to discuss briefly the connection of
the argument of section $2$ to boundary conformal field theory
(BCFT) and SLE. Firstly let's define boundary condition for height
models by following \cite{SB,Cardy ADE}: For the height models on
the square lattice suppose the wired boundary conditions on the real
line but with height $a$ on the left part and $b$ on the right side.
If we work on the half plane then as well as some nested loops we
will have a curve going from the origin to infinity if $a$ and $b$
are adjacent on the corresponding graph, if they don't then it is
possible to have more than one curve. Let's first suppose the
adjacent case then we will have a curve and following \cite{riva
cardy} it is possible to relate the probability that the curve
passes between two closed points to a holomorphic operator with spin
$s$, mentioned in the previous section, with the expectation value
$<\psi (z)>\approx \frac{1}{z^{p}}$. This argument could be the
starting point for a rigorous proof of the connection of these
curves to SLE as Smirnov used it to prove the convergence in the
Ising case \cite{smirnov}. The above definition is possible for all
of the examples that we discussed in section $3$ independent of the
null vector property of the corresponding primary operator. This is
paradoxical because we know from the work of Bauer and Bernard
\cite{Bauer and Bernard1}, and Friedrich and Werner\cite{werner}
that the existence of null vectors at level two is essential to get
SLE. The generalized stochastic equations of
\cite{Gruzberg,Santachiara} showed that the extra symmetries in the
theory can modify the level two null vector to more complicated null
vector, so the Virasoro null vector is not complete for describing
theories with extra symmetries. This can be one of the reasons that
we can get a large set of loop models for different CFTs without
concerning the null vector properties of operators, however the
exact connection is missing so far.

The other important point in the connection of our loop models to
BCFT is: why we are just working with bulk operators without
speaking about boundary states? This comes from the old result of
Cardy \cite{cardybcft} stating that there is a bijection between the
possible conformally invariant boundary conditions and the bulk
primary operators. Let's see this more carefully: in the upper half
plane for the above boundary condition it was shown in
\cite{cardybcft} that only one copy of the chiral algebra acts and
so it is possible to write the partition function on the cylinder as
a sum on the characters of chiral parts
\begin{eqnarray}\label{BCFt}
Z_{ba}=\sum_{i}n_{ib}^{a}\chi_{i}(\tau),
\end{eqnarray}
where sum is over all of the primary operators and $n_{ib}^{a}$ is
the element of matrix $n_{i}$ satisfying Verlinde fusion algebra
\cite{BPPZ} with the eigenvalues the same as for the fusion matrix.
So one can write the graph corresponding to $n_{i}$ and find the
same loop representation as we found. This is the method followed by
\cite{BPPZ} to classify all the conformal boundary conditions of
rational CFTs specially $SU(2)$ WZW models. The above argument
teaches us that the discussed loop models are compatible with
conformal boundary conditions and so with SLE however it seems that
this statement is highly nontrivial and need more investigation
\cite{rajabpour2}. For example however we used the Cardy's equation
as a consistency condition for conformal boundary conditions but it
is not trivial why in the lattice model when we go to the continuum
the model is conformally invariant. This comes from an old problem
in statistical physics at the critical point that: is the
statistical mechanics models at this point conformally invariant or
invariant at least at the critical loop levels? There are many
evidences that the answer is yes for most of the exactly solved
models but SLE is the only method so far giving us a machinery for
rigorous proof.

\section{Conclusions}\

 We showed that it is possible to define a height model for
 a generic primary operator of general CFT by using the fusion matrix as an adjacency matrix. In
 addition it is possible to associate to these height models some
 loop models with loop weights equal to the quantum dimension of
 corresponding primary operator.

 In the critical regimes, these new loop models
 have some properties similar to those
 of loops  corresponding
 to  CFT. For example, the loop model corresponding to $M(3,4)$
 has the same properties as the Ising model's loop representation.  In the lattice level it is not
 completely clear why this should happen but it gives us a good
 machinery to find the loop properties of lattice models by
 studying the corresponding CFT without going to lattice level
 and studying the properties of discrete variables which most
 of the times are quite difficult. Investigating the
 connection with the integrable ADE models is enlightening in
 better understanding the connection with lattice models. Of course finding the CFT of the
 lattice model is not simple and so we think that our method can
 be useful for the more studied models with the known operator
 content. In this paper we just list some simple CFTs but as we
 mentioned before it is possible to investigate the loop models
 corresponding to the more general CFTs from affine Lie algebras
 to supersymmetric models and coset models. It is also interesting to generalize the height models to the fused cases and
 find the connection to the loop models with two kind of loops
 \cite{Fendley2}. The other more interesting direction is using the
 method of Behrend, Pearce, Petkova and
Zuber \cite{BPPZ} to find the fusion rules of boundary conformal
field theory by just using the modular invariant partition function
of the CFT and then finding the loop model representation by the
method that we described in the paper \cite{rajabpour2}.

 Moreover we think that
 this method of generation of loop models can be useful to
 classify the ground state of topological quantum theories
 because as we already  mentioned before the weights of the above loop models are
 exactly the same as quantum dimensions of the operators appearing
 in the CFT. The primary operators of CFT describe the edge
 states of topological theory and so some statistical properties
 of the model specially the topological quantum entanglement
 \cite {kitaev,wen}.
\newline\textbf{Acknowledgments}

I thank John Cardy for careful reading of the manuscript and Paul
Fendley for useful comments and my special thanks goes to Roberto
Tateo for careful reading of the manuscript and stimulating
discussions. \vspace{1cm}
\begin{appendix}

\section{Appendix: Loop Models for More General Coset Theories
}\label{appdisk}\setcounter{equation}{0}\

 In this appendix we are going to summarize some of the possible
 loop  models of more general coset theories.

 $\frac{SU_{k}(2)\oplus SU_{l}(2)}{SU_{k+l}(2)}$: the case $l=1$ was
 already discussed in section ~3. The $l=2$ is
 related to $N=1$ superconformal minimal models and should have
 critical loops related to the loops of $SU_{k}(2)$. To
 get all of the critical loop representations of $N=1$
 superconformal models one should investigate the $S$ matrix of
 these models carefully. Let's see this case with more detail. Take $k=m-2$ then the
 central charge of this model is
 $c=\frac{3}{2}(1-\frac{8}{m(m+2)})$ for $m=3,4,5,...$ and one can
 label the conformal weights by the triplets $(j,k,l)$ with $k=0,1,...,m$, $j=0,1,...,m-2$ and
 $l=0,1,2$. We just consider the triplets with $j-k+l$ being even
 and the symmetry $(j,k,l)=(m-2-j,m-k,2-l)$, so by using the $S$
 matrix given in the \cite{Xu,Carpi} we will have the following
 quantum dimensions
\begin{eqnarray}\label{super}
d_{(j,k,l)}=\frac{\sin(\pi\frac{j+1}{m})\sin(\pi\frac{k+1}{m+2})\sin(\pi\frac{l+1}{4})}{\sin(\frac{\pi}{m+2})\sin(\frac{\pi}{m})\sin(\frac{\pi}{4})}.
\end{eqnarray}

 For $m=3$ this theory is equal to the $M(5,4)$ minimal model
 describing three critical Ising model and one can find the same
 results as the example 1. For $m=4$ one can find the following
 conformal dimensions,

\begin{eqnarray}\label{super2}
d_{(0,0,0)}&=&1,\hspace{1.4cm}d_{(0,2,0)}=2,\hspace{1cm}d_{(0,4,0)}=1,\hspace{1cm}d_{(1,1,0)}=\sqrt{6},\hspace{1cm}\nonumber\\
d_{(1,3,0)}&=&\sqrt{6},\hspace{1cm}d_{(2,0,0)}=1,\hspace{1cm}d_{(2,2,0)}=2,\hspace{1.1cm}d_{(2,4,0)}=1,\hspace{1cm}\nonumber\\
d_{(0,3,1)}&=&\sqrt{6},\hspace{1cm}d_{(1,4,1)}=2,\hspace{1cm}d_{(2,3,1)}=\sqrt{6},\hspace{.85cm}d_{(1,2,1)}=2\sqrt{3}.
\end{eqnarray}
Using equation (\ref{S minimal}) one can find the drift of CLE
easily. The extension to general $m$ shows that we have a primary
operator $(0,m-1,0)$ for odd $m$ with the graph related to $A_{m+1}$
and so with the following CLE drifts
\begin{eqnarray}\label{Z(k2)}
\kappa=4\frac{m+2}{m+3},\hspace{2cm}\kappa=4\frac{m+2}{m+1},
\end{eqnarray}
for dilute and dense cases respectively. If we drop the constraint
on the triplets then we can find the same equation for even $m$ if
we chose $(0,m-1,0)$ which is indeed possible if we work with
modified partition function \cite{Carpi}. It is not difficult to
show that for $(m-3,0,0)$ the corresponding graph is $A_{m-1}$.
These loop models can be related to the paper \cite{Fendley} which
gives a lattice loop candidate for supersymmetric models.

Finally we conjecture that for the more
 general case, $\frac{SU_{k}(2)\oplus SU_{l}(2)}{SU_{k+l}(2)}$, we should
 have at least the following conformal curves,
\begin{eqnarray}\label{coset}
\kappa=4\frac{k+2}{k+3}, \hspace{1cm}\kappa=4\frac{l+2}{l+3},
\hspace{1cm}\kappa=4\frac{k+l+2}{k+l+3}.
\end{eqnarray}
We think that these are related to critical loops of fused RSOS
models introduced in \cite{DJMO}. The important thing to mention
here is most of the loop models that we can extract with our method
in this case are non-critical and can not be consistent. There is a
method for resolving this by using the coupled loop models
introduced recently by Fendley \cite{Fendley}\footnote{I am grateful
 Paul Fendley for mentioning this point to me} which is based on decoupling the quantum dimension to
 the product of generally two different loop weights for fused graphs.

$\frac{SU_{k-1}(N)\oplus SU_{1}(N)}{SU_{k}(N)}$: These coset models
are called $W_{N}$ algebra and describing the $\textbf{A}^{n}$
critical models \cite{de Vega}, the central charge of these models
is $c=(N-1)(1-\frac{N(N+1)}{(k+N)(k+N-1)})$. For these models by
getting motivation from the quantum dimensions that we calculated
for the $SU_{k}(N)$ one can show that for every $N>2$ we have just
critical loop models with the following CLE drift for $k=2$,
\begin{eqnarray}\label{W}
\kappa=4\frac{N+2}{N+3}.
\end{eqnarray}
It is not obvious that why the $k=2$ case is an exception here but
this formalism is showing that for $k> 2$ we can not find a
nontrivial critical loop model. It seems that there should be also
the same decoupling procedure as \cite{Fendley} however this case is
totally unexplored. The above formula is giving the true answer for
$N=3$ and $k=2$ which is related to three states Potts model. It is
not difficult to see that $k=2$ is related to $Z(N)$ model and the
above equation is in fact the same as the equation (\ref{Z(k)}) that
we found for simplest parafermionic model.

\end{appendix}

\end{document}